\documentclass[twoside]{article}%
\usepackage{amssymb}
\usepackage{amsfonts}
\usepackage{amsmath}
\usepackage{graphicx}%
\providecommand{\U}[1]{\protect\rule{.1in}{.1in}}
\newtheorem{theorem}{Theorem}

\newenvironment{proof}[1][Proof]{\noindent\textbf{#1.} }{\ \rule{0.5em}{0.5em}}
\begin{document}

\title{\textbf{Blowup for the Euler and Euler-Poisson Equations with Repulsive Forces
II}}
\author{M\textsc{anwai Yuen\thanks{E-mail address: nevetsyuen@hotmail.com }}\\\textit{Department of Applied Mathematics,}\\\textit{The Hong Kong Polytechnic University,}\\\textit{Hung Hom, Kowloon, Hong Kong}}
\date{Revised 23-Dec-2010}
\maketitle

\begin{abstract}
In this paper, we continue to study the blowup problem of the $N$-dimensional
compressible Euler or Euler-Poisson equations with repulsive forces, in radial
symmetry. In details, we extend the recent result of "M.W. Yuen,
\textit{Blowup for the Euler and Euler-Poisson Equations with Repulsive
Forces}, Nonlinear Analysis Series A: Theory, Methods \& Applications
\textbf{74} (2011), 1465--1470.". We could further apply the integration
method to obtain the more general results which the non-trivial classical
solutions $(\rho,V)$, with compact support in $[0,R]$, where $R>0$ is a
positive constant with $\rho(t,r)=0$ and $V(t,r)=0$ for $r\geq R$, under the
initial condition%
\begin{equation}
H_{0}=\int_{0}^{R}r^{n}V_{0}dr>0
\end{equation}
where an arbitrary constant $n>0$, blow up on or before the finite time
$T=2R^{n+2}/(n(n+1)H_{0})$ for pressureless fluids or $\gamma>1.$ The results
obtained here fully cover the previous known case for $n=1$.



\end{abstract}

\section{Introduction}

The compressible isentropic Euler $(\delta=0)$ or Euler-Poisson $(\delta
=\pm1)$ equations can be written in the following form:
\begin{equation}
\left\{
\begin{array}
[c]{rl}%
{\normalsize \rho}_{t}{\normalsize +\nabla\cdot(\rho u)} & {\normalsize =}%
{\normalsize 0}\\
\rho\lbrack u_{t}+(u\cdot\nabla)u]{\normalsize +\nabla}P & {\normalsize =}%
{\normalsize \rho\nabla\Phi}\\
{\normalsize \Delta\Phi(t,x)} & {\normalsize =\delta\alpha(N)}%
{\normalsize \rho}%
\end{array}
\right.  \label{Euler-Poisson}%
\end{equation}
where $\alpha(N)$ is a constant related to the unit ball in $R^{N}$. As usual,
$\rho=\rho(t,x)\geq0$ and $u=u(t,x)\in\mathbf{R}^{N}$ are the density and the
velocity respectively. $P=P(\rho)$\ is the pressure function. The $\gamma$-law
for the pressure term $P(\rho)$ could be applied:%
\begin{equation}
{\normalsize P}\left(  \rho\right)  {\normalsize =K\rho}^{\gamma}
\label{gamma}%
\end{equation}
which the constant $\gamma\geq1$. If\ ${\normalsize K>0}$, we call the system
with pressure; if ${\normalsize K=0}$, we call it pressureless.\newline When
$\delta=-1$, the system is self-attractive. The equations (\ref{Euler-Poisson}%
) are the Newtonian descriptions of a galaxy in astrophysics \cite{BT} and
\cite{C}. When $\delta=1$, the system is the compressible Euler-Poisson
equations with repulsive forces. It can be used as a semiconductor model
\cite{Cse}. For the compressible Euler equation with $\delta=0$, it is a
standard model in fluid mechanics \cite{Lions}. And the Poisson equation
(\ref{Euler-Poisson})$_{3}$ could be solved by%
\begin{equation}
{\normalsize \Phi(t,x)=\delta}\int_{R^{N}}G(x-y)\rho(t,y){\normalsize dy}%
\end{equation}
where $G$ is Green's function:
\begin{equation}
G(x)\doteq\left\{
\begin{array}
[c]{ll}%
|x|, & N=1\\
\log|x|, & N=2\\
\frac{-1}{|x|^{N-2}}, & N\geq3.
\end{array}
\right.
\end{equation}

For the construction of analytical solutions for the systems, interested
readers could refer to \cite{GW}, \cite{M1}, \cite{DXY}, \cite{Li} and
\cite{Y1}. The local existence for the systems can be found in \cite{Lions},
\cite{M2}, \cite{B} and \cite{G}. The analysis of stabilities for the systems
may be referred to \cite{SI}, \cite{A}, \cite{E}, \cite{MUK}, \cite{MP},
\cite{P}, \cite{DLY}, \cite{DXY}, \cite{Y2}, \cite{CT} and \cite{CH}.

The solutions in radial symmetry could be in this form:
\begin{equation}
\rho=\rho(t,r)\text{ and }u=\frac{x}{r}V(t,r)=:\frac{x}{r}V
\end{equation}
with the radius $r=\left(  \sum_{i=1}^{N}x_{i}^{2}\right)  ^{1/2}$.

The Poisson equation (\ref{Euler-Poisson})$_{3}$ becomes%
\begin{equation}
{\normalsize r^{N-1}\Phi}_{rr}\left(  {\normalsize t,x}\right)  +\left(
N-1\right)  r^{N-2}\Phi_{r}{\normalsize =}\alpha\left(  N\right)
\delta{\normalsize \rho r^{N-1}}%
\end{equation}%
\begin{equation}
\Phi_{r}=\frac{\alpha\left(  N\right)  \delta}{r^{N-1}}\int_{0}^{r}%
\rho(t,s)s^{N-1}ds.
\end{equation}
By standard computation, the systems in radial symmetry can be rewritten in
the following form:%
\begin{equation}
\left\{
\begin{array}
[c]{c}%
\rho_{t}+V\rho_{r}+\rho V_{r}+\dfrac{N-1}{r}\rho V=0\\
\rho\left(  V_{t}+VV_{r}\right)  +P_{r}(\rho)=\rho\Phi_{r}\left(  \rho\right)
.
\end{array}
\right.  \label{eq12345}%
\end{equation}

In literature, Makino, Ukai and Kawashima first studied the blowup of tame
solutions \cite{MUK} for the compressible Euler equations $(\delta=0)$. Then
Makino and Perthame investigated the corresponding tame solutions for the
system with gravitational forces \cite{MP}. After that Perthame \cite{P}
obtained the blowup results for the $3$-dimensional pressureless system with
repulsive forces $(\delta=1)$. There are other blowup results for the systems
in \cite{E}, \cite{ELT}, \cite{CT} and \cite{CH}.

Very recently, Yuen \cite{YuenNA} used the integration method to show that
with the initial velocity
\begin{equation}
H_{0}=\int_{0}^{R}rV_{0}dr>0, \label{condition1}%
\end{equation}
the solutions with compact support to the Euler $(\delta=0)$ or Euler-Poisson
equations with repulsive forces $(\delta=1)$ blow up in the finite time.

In this article, we can observe that the condition (\ref{condition1}) could be
more general to have the corresponding blowup results. In fact, we could
further apply the integration method to extend Yuen's result as the following theorem:

\begin{theorem}
\label{thm:1 copy(1)}Consider the Euler $(\delta=0)$ or Euler-Poisson
equations with repulsive forces $(\delta=1)$ (\ref{Euler-Poisson}) in $R^{N}$.
The non-trivial classical solutions $\left(  \rho,V\right)  $, in radial
symmetry, with compact support in $\left[  0,R\right]  $, where $R>0$ is a
positive constant ($\rho(t,r)=0$ and $V(t,r)=0$ for $r\geq R$) and the initial
velocity:
\begin{equation}
H_{0}=\int_{0}^{R}r^{n}V_{0}dr>0
\end{equation}
with an arbitrary constant $n>0,$\newline blow up on or before the finite time
$T=2R^{n+2}/(n(n+1)H_{0}),$ for pressureless fluids $(K=0)$ or $\gamma>1$.
\end{theorem}

We remark that the condition
\begin{equation}
\rho(t,r)=0\text{ and }V(t,r)=0\text{ for }r\geq R
\end{equation}
in the theorem is called non-slip boundary condition \cite{Day} and \cite{CTC}.

\section{Integration Method}

We just follow the integration method which was designed in \cite{YuenNA} to
obtain the further results.

\begin{proof}
The density function $\rho(t,x(t;x))$ preserves its non-negative nature as we
can integrate the mass equation (\ref{Euler-Poisson})$_{1}$:%
\begin{equation}
\frac{D\rho}{Dt}+\rho\nabla\cdot u=0 \label{eqq2}%
\end{equation}
with the material derivative,%
\begin{equation}
\frac{D}{Dt}=\frac{\partial}{\partial t}+\left(  u\cdot\nabla\right)
\label{eqq1}%
\end{equation}
to have:%
\begin{equation}
\rho(t,x)=\rho_{0}(x_{0}(0,x_{0}))\exp\left(  -\int_{0}^{t}\nabla\cdot
u(t,x(t;0,x_{0}))dt\right)  \geq0
\end{equation}
for $\rho_{0}(x_{0}(0,x_{0}))\geq0,$ along the characteristic curve.

Then we can manipulate the momentum equation (\ref{eq12345})$_{2}$ for the
non-trivial solutions in radial symmetry, $\rho_{0}\neq0$, to obtain:%
\begin{equation}
V_{t}+VV_{r}+K\gamma\rho^{\gamma-2}\rho_{r}=\Phi_{r}%
\end{equation}%
\begin{equation}
V_{t}+\frac{\partial}{\partial r}(\frac{1}{2}V^{2})+K\gamma\rho^{\gamma-2}%
\rho_{r}=\Phi_{r} \label{eq478}%
\end{equation}%
\begin{equation}
r^{n}V_{t}+r^{n}\frac{\partial}{\partial r}(\frac{1}{2}V^{2})+K\gamma
r^{n}\rho^{\gamma-2}\rho_{r}=r^{n}\Phi_{r} \label{eq789}%
\end{equation}
with multiplying the more general function $r^{n}$ with $n>0$, on the both
sides.\newline We notice that this is the critical step in this paper to
extend the blowup result in \cite{YuenNA}.\newline We could take the
integration with respect to $r,$ to equation (\ref{eq789}), for $\gamma>1$ or
$K\geq0$:%
\begin{equation}
\int_{0}^{R}r^{n}V_{t}dr+\int_{0}^{R}r^{n}\frac{d}{dr}(\frac{1}{2}V^{2}%
)+\int_{0}^{R}K\gamma r^{n}\rho^{\gamma-2}\rho_{r}dr=\int_{0}^{R}r^{n}\Phi
_{r}dr
\end{equation}%
\begin{equation}
\int_{0}^{R}r^{n}V_{t}dr+\int_{0}^{R}r^{n}\frac{d}{dr}(\frac{1}{2}V^{2}%
)+\int_{0}^{R}\frac{K\gamma r^{n}}{\gamma-1}d\rho^{\gamma-1}=\int_{0}%
^{R}\left[  \frac{\alpha(N)\delta r^{n}}{r^{N-1}}\int_{0}^{r}\rho
(t,s)s^{N-1}ds\right]  dr
\end{equation}%
\begin{equation}
\int_{0}^{R}r^{n}V_{t}dr+\int_{0}^{R}r^{n}\frac{d}{dr}(\frac{1}{2}V^{2}%
)+\int_{0}^{R}\frac{K\gamma r^{n}}{\gamma-1}d\rho^{\gamma-1}\geq0
\label{eq567}%
\end{equation}
for $\delta\geq0.$\newline Then, the below equation can be showed by
integration by part:%
\begin{equation}%
\begin{array}
[c]{c}%
\int_{0}^{R}r^{n}V_{t}dr-\frac{1}{2}\int_{0}^{R}nr^{n-1}V^{2}dr+\frac{1}%
{2}\left[  R^{n}V^{2}(t,R)-0^{n}\cdot V^{2}(t,0)\right] \\
-\int_{0}^{R}\frac{K\gamma nr^{n-1}}{\gamma-1}\rho^{\gamma-1}dr+\frac{K\gamma
}{\gamma-1}\left[  R^{n}\rho^{\gamma-1}(t,R)-0^{n}\cdot\rho^{\gamma
-1}(t,0)\right]  \geq0.
\end{array}
\end{equation}
The above inequality with the boundary condition ($V(t,R)=0$ and $\rho
(t,R)=0$), becomes%
\begin{equation}
\int_{0}^{R}r^{n}V_{t}dr-\frac{1}{2}\int_{0}^{R}nr^{n-1}V^{2}dr-\int_{0}%
^{R}\frac{K\gamma nr^{n-1}}{\gamma-1}\rho^{\gamma-1}dr\geq0
\end{equation}%
\begin{equation}
\frac{d}{dt}\int_{0}^{R}r^{n}Vdr-\frac{1}{2}\int_{0}^{R}nr^{n-1}V^{2}%
dr-\int_{0}^{R}\frac{K\gamma nr^{n-1}}{\gamma-1}\rho^{\gamma-1}dr\geq0
\end{equation}%
\begin{equation}
\frac{d}{dt}\frac{1}{n+1}\int_{0}^{R}Vdr^{n+1}-\frac{1}{2}\int_{0}^{R}\frac
{n}{(n+1)r}V^{2}dr^{n+1}\geq\int_{0}^{R}\frac{K\gamma nr^{n-1}}{\gamma-1}%
\rho^{\gamma-1}dr\geq0
\end{equation}
for $n>0$ and $\gamma>1$ or $K=0.$\newline For non-trivial initial density
functions $\rho_{0}\geq0$, we obtain:%
\begin{equation}
\frac{d}{dt}\frac{1}{n+1}\int_{0}^{R}Vdr^{n+1}-\frac{1}{2}\int_{0}^{R}\frac
{n}{(n+1)r}V^{2}dr^{n+1}\geq0
\end{equation}%
\begin{equation}
\frac{d}{dt}\frac{1}{n+1}\int_{0}^{R}Vdr^{n+1}\geq\int_{0}^{R}\frac
{n}{2(n+1)r}V^{2}dr^{n+1}\geq\frac{n}{2(n+1)R}\int_{0}^{R}V^{2}dr^{n+1}
\label{eq11111}%
\end{equation}%
\begin{equation}
\frac{d}{dt}\int_{0}^{R}Vdr^{n+1}\geq\frac{n}{2R}\int_{0}^{R}V^{2}dr^{n+1}.
\label{eq2222}%
\end{equation}
We denote
\begin{equation}
H:=H(t)=\int_{0}^{R}r^{n}Vdr=\frac{1}{n+1}\int_{0}^{R}Vdr^{n+1}%
\end{equation}
and apply the Cauchy-Schwarz inequality:%
\begin{equation}
\left\vert \int_{0}^{R}V\cdot1dr^{n+1}\right\vert \leq\left(  \int_{0}%
^{R}V^{2}dr^{n+1}\right)  ^{1/2}\left(  \int_{0}^{R}1dr^{n+1}\right)  ^{1/2}%
\end{equation}%
\begin{equation}
\left\vert \int_{0}^{R}V\cdot1dr^{n+1}\right\vert \leq\left(  \int_{0}%
^{R}V^{2}dr^{n+1}\right)  ^{1/2}\left(  R^{n+1}\right)  ^{1/2}%
\end{equation}%
\begin{equation}
\frac{\left\vert \int_{0}^{R}Vdr^{n+1}\right\vert }{R^{\frac{n+1}{2}}}%
\leq\left(  \int_{0}^{R}V^{2}dr^{n+1}\right)  ^{1/2}%
\end{equation}%
\begin{equation}
\frac{(n+1)^{2}H^{2}}{R^{n+1}}\leq\int_{0}^{R}V^{2}dr^{n+1}%
\end{equation}%
\begin{equation}
\frac{n(n+1)^{2}H^{2}}{2R^{n+2}}\leq\frac{n}{2R}\int_{0}^{R}V^{2}dr^{n+1}
\label{ineq123}%
\end{equation}
for driving equation (\ref{eq2222}) to be%
\begin{equation}
\frac{d}{dt}(n+1)H\geq\frac{n}{2R}\int_{0}^{R}V^{2}dr^{n+1}\geq\frac
{n(n+1)^{2}H^{2}}{2R^{n+2}}%
\end{equation}
with inequality (\ref{ineq123}),%
\begin{equation}
\frac{d}{dt}H\geq\frac{n(n+1)H^{2}}{2R^{n+2}}%
\end{equation}%
\begin{equation}
H\geq\frac{-2R^{n+2}H_{0}}{n(n+1)H_{0}t-2R^{n+2}}.
\end{equation}
Finally, we could require the initial condition
\begin{equation}
H_{0}=\int_{0}^{R}r^{n}V_{0}dr>0
\end{equation}
for showing that the solutions blow up on or before the finite time
$T=2R^{n+2}/(n(n+1)H_{0}).$

This completes the proof.
\end{proof}

We notice that the results in this paper fully cover the previous case for
$n=1$ \cite{YuenNA}. Further researches are needed to have the corresponding
results for the non-radial symmetric cases.

\end{document}